\documentclass[namedreferences]{kluwer}
\usepackage{graphicx}
\begin{document}
\begin{article}
\begin{opening}
\title{Close binary EHB stars from SPY%
\footnotemark[1]\footnotemark[2]\footnotemark[3]
\protect\footnotetext[1]{Based on data obtained at the
    Paranal Observatory of the European Southern Observatory for programs
    165.H-0588, 167.D-0407, 70.D-0334(A), 71.D-0383(A)}
\protect\footnotetext[2]{Based on observations collected at the
    German-Spanish Astronomical Center, Calar Alto, operated by the
    Max-Planck-Institut f\"ur Astronomie Heidelberg jointly with the Spanish
    National Commission for Astronomy}
\protect\footnotetext[3]{Based on data obtained with
    the WHT of the Isaac Newton Telescope operated on the island of La Palma
    by the Isaac Newton Group in the Spanish Observatory del Roque de los
    Muchachos of the Instituto de Astrofisica de Canarias}
}            

\author{R. \surname{Napiwotzki}}%\email{napiwotzki@sternwarte.uni-erlangen.de}}
\author{C.A. \surname{Karl}}
\author{T. \surname{Lisker}} 
\author{U. \surname{Heber}}
\institute{Dr.~Remeis-Sternwarte, Universit\"at Erlangen-N\"urnberg, 
    Bamberg, Germany}
\author{N. \surname{Christlieb}}
\author{D. \surname{Reimers}}
\institute{Hamburger Sternwarte, Universit\"at Hamburg, Hamburg, Germany}
\author{G. \surname{Nelemans}}
\institute{Institute of Astronomy, University of Cambridge, Cambridge, UK}
\author{D. \surname{Homeier}}
\institute{Dept.\ of Physics \& Astronomy, University of Georgia, Athens, USA}
%\date: rather not

\runningtitle{Close binary EHB stars from SPY}
\runningauthor{R.~Napiwotzki et al.}
 
\begin{abstract} 
  We present the results of a radial velocity (RV) survey of 46 subdwarf B
  (sdB) and 23 helium-rich subdwarf O (He-sdO) stars. We detected 18 (39\%)
  new sdB binary systems, but only one (4\%) He-sdO binary. Orbital parameters
  of nine sdB and sdO binaries, derived from follow-up spectroscopy, are
  presented.  Our results
  are compared with evolutionary scenarios and previous observational
  investigations.
\end{abstract}


% \abbreviations{\abbrev{KAP}{Kluwer Academic Publishers};
%    \abbrev{compuscript}{Electronically submitted article}}

% \nomenclature{\nomen{KAP}{Kluwer Academic Publishers};
%    \nomen{compuscript}{Electronically submitted article}}

% \classification{JEL codes}{D24, L60, 047}
\end{opening}

\section{Introduction}
In the standard picture (e.g.\ \opencite{Heb86}) subdwarf B (sdB) stars are
core helium burning stars on the extended horizontal branch (EHB). They
consist of a He core with a mass of $\approx$0.48$M_\odot$, the canonical mass
for a He core flash on the first red giant branch (RGB), and a thin, inert
hydrogen shell. Evolutionary models of EHB stars calculated by
\inlinecite{DRO93} and \inlinecite{DDR96} adopted an enhanced mass loss on the
RGB, however, without specifying a particular mechanism.

Recent radial velocity (RV) surveys revealed that a large fraction of all
sdB stars resides in close binaries (\opencite{SLY98}; 
\opencite{MHM01}; \opencite{EHL03}). \citeauthor{HPM02} (\citeyear{HPM02},
\citeyear{HPM03}) performed a theoretical investigation of possible close
binary channels 
 for the formation of sdBs: stable Roche lobe overflow, common envelope 
ejection, and merging of two He core white dwarf (WDs).  
They concluded that it is possible to
explain {\em all} sdB stars as the result of close binary evolution.

The helium-rich subdwarf O stars (He-sdO) are hotter than sdBs, but have
similar surface gravities. Their formation is still a mystery. In one scenario
they are explained as an advanced stage of post-EHB evolution
(\opencite{WWC82}, \opencite{DHW90}). Another scenario proposes the formation 
of He-sdOs via a phase as extreme He stars after the merging of two He core 
WDs (cf.\ discussion in \opencite{AJ03}).

SPY, the ESO {\bf S}upernovae type Ia {\bf P}rogenitor surve{\bf Y}
(\citeauthor{NCD01} \citeyear{NCD01}, \citeyear{NCD03}), is a
programme dedicated to search for short period binary WDs (double
degenerates -- DDs). The aim of SPY is the detection of DD progenitors of
supernovae type Ia by means of a survey for RV variations. 
SN\,Ia progenitor candidates should be close enough to
merge within one Hubble time due to gravitational wave radiation and the
combined mass should exceed the Chandrasekhar limit for WDs.  
The SPY input catalogue collected WD candidates brighter than
$B=16.5$ from a variety of source catalogues (cf.\ \opencite{NCD01}). 
The sdBs analyzed in this paper were selected from 
the Hamburg/ESO survey
\cite{WKG96}, the \inlinecite{MS99} WD catalogue, 
and the Hamburg Quasar
Survey \cite{HGE95}. 

As a by-product, SPY produced
high accuracy RV measurements of more than 46 sdB stars and 23 He-sdO stars,
which were included in our sample because of misclassifications in the input
catalogues.
Thus SPY produces a large sample of known close binary sdB stars,
offering an independent data set and 
allowing for a substantial improvement of the statistics. The He-sdO
sample included in SPY is the first He-sdO sample systematically checked for
RV variations. This allows us to investigate the role of binarity for their
formation and their possible link with the sdB stars. 

\section{Results of the  radial velocity survey}
SPY is carried out with the high-resolution spectrograph UVES at the UT2
telescope of ESO VLT. With our instrument setup we achieve nearly complete
spectral coverage from 3200\,\AA\ to 6650\,\AA with only two $\approx$80\,\AA\ 
wide gaps. Since SPY was implemented as a bad weather program, we used a wide
($2.1''$) slit to minimize slit losses. The resulting spectral resolving power
is $R=18\,500$ (0.36\,\AA\ at H$\alpha$) or better, if seeing disks were
smaller than the slit width. Due to the nature of the project, two spectra at
different ``random'' epochs separated by at least one day are observed (cf.\ 
\opencite{NCD01}, for details). The RV variations were measured differentially
between both observed spectra with a cross correlation routine. This allowed
to include all suitable lines (H, He, and metal lines) for the RV
determination in a very flexible way.  We routinely achieved a RV accuracy of
2\,km/s or better.

\subsection{The sdB sample}
\label{s:sdbsample}

\begin{figure}
\centerline{\includegraphics[width=7.5cm,angle=-90]{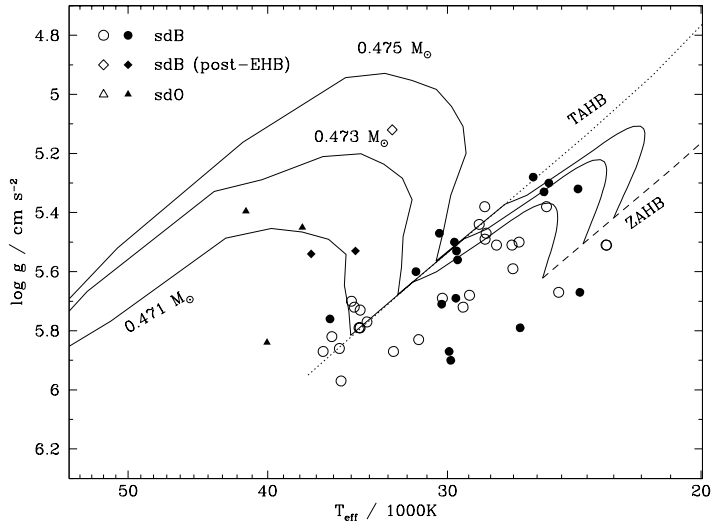}}
\caption{The sample of sdBs checked for RV variations in the 
  temperature-gravity diagram (parameters from \protect\opencite{LHN03b}).
  Filled symbols indicate RV variable binaries. The sdBs indicates as
  ``post-EHB" are rejected by the ``strip-selection'' discussed in the text.
  Representative evolutionary tracks from \protect\inlinecite{DRO93} and the 
zero-age (ZAHB) and terminal-age (TAHB) horizontal branches are indicated.}
\label{f:hr}
\end{figure}

\begin{figure}
\centerline{\includegraphics[width=7.5cm,angle=-90]{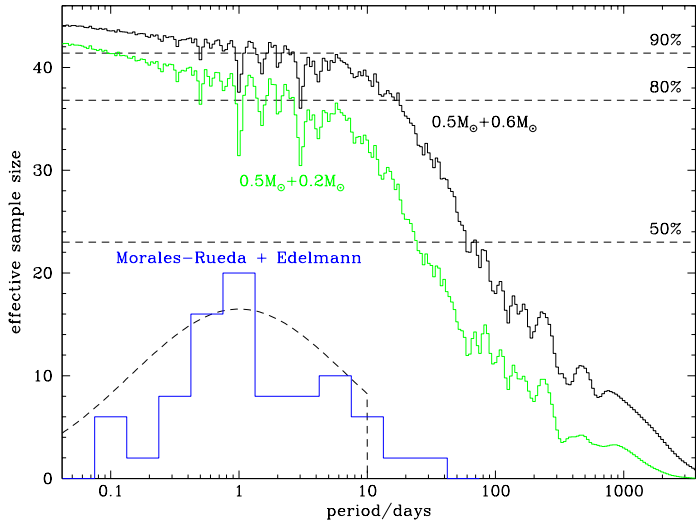}}
\caption{Detection efficiency as a function of orbital period for two
different assumptions on the companion mass. For comparison we display the
period distribution of sdBs for the combined sample of 
\protect\inlinecite{MMM03} and
\protect\inlinecite{EHL03}. The dashed line indicates the period distribution
adopted by \protect\inlinecite{MHM01} for estimating their detection 
efficiency.}
\label{f:detection}
\end{figure}

RV variations could be checked for 46 single-lined sdBs from the complete
SPY sample (\opencite{LHN03a}, \citeyear{LHN03b}).
Their distribution in the
temperature-gravity plane is displayed in Fig.~\ref{f:hr}. Our RV measurements
yielded 18 RV variable binaries, i.e.\ 39\% of this sample. This indicates a 
high fraction of close binaries in the SPY sample, conforming the importance
of binary channels for the formation of sdBs. However, somewhat surprisingly
our result points to a lower fraction of detections than in the
\inlinecite{MHM01} sample, who found 58\% RV variable sdBs. 

For a quantitative comparison with other samples and an evaluation of the
true fraction of binaries one needs to know the detection efficiency, i.e.\ 
the chance that a binary escaped
detection, because of unfavorable inclination angles or phasing
of the observations. The detection efficiency obviously depends on the orbital
period and the mass of the unseen companion. We performed a Monte Carlo
simulation of our observed sample for two different choices of the mass of the
invisible companion: $0.6M_\odot$, the mass of a
typical WD, and $0.2M_\odot$ for a typical low mass main sequence
star. For every star we calculated detection probabilities taking into
account the timing of exposures and the RV accuracy. Results for the sample
were coadded yielding the effective sample size (i.e.\ the
size of an equivalent sample with 100\% binary detection rate) plotted in 
Fig.~\ref{f:detection} as function
of orbital period. For an alternative interpretation we indicate the relative
percentage of the detection rate. The detection probabilities
are quite high for short period systems, exceeding 90\% for systems with a
WD companion and periods shorter than one day. 

\begin{table}
\caption{Corrections for detection efficiency for the sdB and the He-sdO 
  samples. Calculations
  were done for the two distributions discussed in the text, the canonical sdB
  mass of $0.5M_\odot$, and different companion masses.}
\label{t:det-sdB}
\label{t:det-sdO}
\begin{tabular}{llllll} \hline
Sample &$M_{\mathrm{comp}}/M_\odot$ & distribution 
&observed  &efficiency &corrected \\
&                          && rate                  &&rate\\ \hline
sdB &0.2    &Gauss &39\%  &88\%   &44\%  \\
&0.2    &flat  &39\%  &87\%   &45\%  \\ \hline
&0.5    &Gauss &39\%  &93\%   &42\%  \\ \hline
&0.6    &Gauss &39\%  &94\%   &42\%  \\ 
&0.6    &flat &39\%  &93\%   &42\%  \\ \hline \hline
He-sdO &0.2    &Gauss &4\%  &89\%   &5\%  \\
& 0.2    &flat  &4\%  &88\%   &5\%  \\ \hline
& 0.6    &Gauss &4\%  &94\%   &5\%  \\ 
& 0.6    &flat &4\%  &94\%   &5\%  \\ \hline
\end{tabular}
\end{table}

However, to estimate the total numbers of binaries in the observed sample, one
has to adopt a period distribution of the underlying binary population. We
corrected the observed number of binaries for two different assumptions: 1) a
truncated Gaussian centered on $P = 1$d
(see Fig.~\ref{f:detection}) 
and 2) a flat distribution
covering the interval from $P = 0.1\,\mathrm{d}\ldots 30$\,d. 
The truncated Gaussian distribution corresponds to one of the
distributions discussed by \inlinecite{MHM01} and agrees reasonably well with
the current sample of sdBs with known orbital periods
(Fig.~\ref{f:detection}). The flat distribution was chosen as some sort of
worst case, because it gives high weight to the long period systems with
relatively bad detection probabilities. 
 
Results are detailed in Table~\ref{t:det-sdB}. The corrections are small and
even in the worst case the likely binary fraction amounts to 45\%,
i.e. 3 undetected systems are predicted. The entry for a companion mass of
$0.5M_\odot$ corresponds to corrections calculated by \inlinecite{MHM01}.
Their corrections are small as well and they estimate a binary fraction of
69\%. If we compute the corrections for our sample in exactly the same way we
derive 42\%. The difference between both samples is statistically
significant. 

However, selection criteria of both samples are slightly
different. \citeauthor{MHM01} excluded stars which lie significantly
above the EHB from their sample (``strip-selection''). This criterion
would exclude the three stars marked as ``post-EHB'' in
Fig.~\ref{f:hr} from our sample. Two of these three are binaries,
which lowers the (corrected) binary frequency to 40\%. We
excluded sdBs with spectral contamination from a  main
sequence companion, because these are usually not found in close binary
systems, while \citeauthor{MHM01} included two known sdBs with composite
spectra.  Moreover, our high resolution UVES spectra allow
a more sensitive detection of cool companions
which would have probably escaped detection in the lower
resolution spectra of \citeauthor{MHM01}. 
Including some or all composite systems
into our sample would lower the frequency of close binaries in the SPY sample
even further. 

\subsection{The He-sdO sample}
\label{s:hesdo}

A total of 23 single-lined He-sdOs was checked for RV variations We detected
{\em only one} RV variable close binary in this sample, and this is a peculiar
object: a double-lined system apparently consisting of two subdwarfs
\cite{LHN03a}.  The detection efficiency is very similar to the value of our
sdB sample (see Table~\ref{t:det-sdO}). The corrected close binary fraction
is 5\%, which is much lower than the percentage observed in any sdB sample,
providing evidence that He-sdOs are not the progeny of sdBs.

\subsection{Individual binary systems}

Detection of RV variations is only a start. Follow-up observations are
necessary to derive important system parameters like orbital periods and
amplitudes, which allow us to compute lower limits on the mass of the unseen
companion via the mass function. This information will allow us to further
constrain scenarios for the formation of sdBs. 

We performed follow-up observations of detected sdB binaries at the 3.5\,m
telescope at Calar Alto, the William Herschel telescope at Roque de los
Muchachos, and the
ESO-VLT. Sufficient data for an unambiguous orbital solution is available for
six sdB stars (including HE\,1047-0436; \opencite{NEH01}). Details are given
in Table~\ref{t:orbits}.

% \begin{figure}
% \centerline{\includegraphics[width=8cm,angle=-90]{pmkeele.ps}}
% \caption{Periods vs. system masses for the sdB and sdO binaries of
%   Table~\protect\ref{t:orbits}. The masses of the unseen companion are
%   estimated for the expected average inclination angle ($i = 52^\circ$).  The
%   Chandrasekhar mass and the limit for merging within a Hubble time are
%   indicated.}
% \label{f:pm}
% \end{figure}

Table~\ref{t:orbits} contains data for three objects which are classified
sdO. Unlike the He-sdOs discussed in
Sect.~\ref{s:hesdo} they are H-rich rather than He-rich
with a spectral
appearance not very different from the sdBs. 
A comparison with the parameter
range covered by the SPY sdBs (Fig.~\ref{f:hr}) demonstrates that they
represent nothing else than the hot end of the sdB distribution.

\section{Discussion and conclusions}

We have presented a RV survey of 46 sdBs and 23 He-sdOs for close
binaries. We detected 18 (39\%) sdB and 1 (4\%) He-sdO binaries,
respectively. Although we qualitatively confirm the results of
\inlinecite{SLY98} and \inlinecite {MHM01}, which indicate a high
binary frequency among sdB stars, our quantitative results are at
variance with \inlinecite{MHM01}, who detected 21 close binaries in a
sample of 36 sdBs, i.e.\ 58\%. Differences in detection efficiency
and selection effects cannot explain the discrepancy, as discussed in
Sect.~\ref{s:sdbsample}. After correction for undetected binaries with
identical assumptions the estimated
true rate of close binaries are 42\% in the SPY sample vs.\ 69\% in the
\inlinecite{MHM01} sample. 

Stars from the SPY sample are typically
much fainter than from the \citeauthor{MHM01} sample, who
concentrated on the brightest known sdB stars. Consequently 2/3 of the sdBs
from the SPY sample are more than 1\,kpc away from the Galactic
plane, while only two of the \citeauthor{MHM01} stars are. Thus we expect
that a significant fraction of the SPY sdBs belong to the thick disk
or the halo populations, which are both old and metal poor,
while most of the \citeauthor{MHM01} sdBs are probably members of the
thin disk. Determination of population membership by a kinematical study of 
both samples would help to verify this explanation. 

\begin{table} 
\caption{Orbital parameters of sdB and sdO stars from SPY. 
We calculated the minimum
  mass $M_2(min)$ from the mass function
and the most probable mass  for 
 $i=52^\circ$.  A primary mass of $0.5M_\odot$ was adopted.}
\label{t:orbits}
\begin{tabular}{lllll} \hline
Object &type &$P$/d  &$M_2(\mathrm{min})/M_\odot$  
&$M_2(i=52^\circ)/M_\odot$  \\ \hline
WD\,0048$-$202    &sdB &7.45    &0.33   &0.47\\
HE\,0532$-$4503   &sdB &0.2656  &0.26   &0.35 \\
HE\,0929$-$0424   &sdB &0.4400  &0.38   &0.54\\
HE\,1047$-$0436   &sdB &1.2133  &0.44   &0.71\\
HE\,2135$-$3749   &sdB &0.9241  &0.35   &0.50\\
HE\,2150$-$0238     &sdB &1.322   &0.50   &0.73\\ \hline
HE\,1059$-$2735   &sdO &0.556   &0.31   &0.42\\ 
HE\,1115$-$0631   &sdO   &5.87    &0.52   &0.76\\
HE\,1318$-$2111   &sdO &0.487   &0.34    &0.48\\ \hline
\end{tabular} 
\end{table}

The very low number of close binaries in our He-sdO sample provides evidence
that He-sdOs, as a class, are not successors of EHB stars. Formation of
He-sdOs by merging of two He core WDs would be in agreement with our
finding.

How do our sdB results compare with the binary population synthesis of
\inlinecite{HPM03}? \citeauthor{HPM03}
 produced several simulated samples of sdBs
resulting from their binary evolution channels with different
parameter choices and presented a best fit sample (their simulation set~2). 
They could
reproduce the properties of observed sdB binaries (from
\opencite{MMM03}) and the distribution in the temperature-gravity
plane (cf.\ \opencite{LHN03a}, \citeyear{LHN03b}). 
Observational selection effects against sdBs with visible
cool companions (GK selection) and against stars above the
EHB (strip selection) were simulated. They predict an observable 
binary frequency for their best
fit model of 45\% (the remainder resulting from merging),
which compares well to our 40\%, if we apply the same criteria to
the SPY sample. 

However, two problems remain. \inlinecite{HPM03} simulated a population with
thin disk characteristics, which should be representative for the
 \inlinecite{MHM01} sample, but this sample yielded a binary frequency of
69\%. On the other hand \inlinecite{HPM03} predict a low binary frequency for
stars above the EHB, because this region should contain many sdBs resulting
from merging. Although a systematic observational investigation of stars in
this region has still to be done, it appears that the binary frequency in this
region is higher than expected (see Fig.~\ref{f:hr}) from the theoretical
simulation. A further complication might arise if some or all of the mergers
produce He-sdOs, instead.
In summary the new theoretical simulations represent the properties of the
observed sdB sample well, but some unsolved problems remain. Future
investigations will help to decide, if these can be solved by minor adjustments
or imply larger revisions.

\newcommand{\pipitem}[4]{\bibitem[\protect\citeauthoryear{#1}{#2}]{#3}#4}
\newcommand{\mnras}{{\it MNRAS}}
\newcommand{\aua}{{\it A\&A}}
\newcommand{\auas}{{\it A\&AS}}
\newcommand{\apj}{{\it ApJ}}
\newcommand{\apjs}{{\it ApJS}}
\newcommand{\an}{{\it AN}}
\newcommand{\apass}{{\it Ap\&SS}}

\end{article}
\end{document}